# Around-the-world seismic echo as a trigger for aftershocks and the main shock of an earthquake

A.V. Guglielmi[1], O.D. Zotov[1,2]

[1] *Schmidt Institute of Physics of the Earth, Russian Academy of Sciences, Moscow, Russia, guglielmi@mail.ru*
[2] *Borok Geophysical Observatory, the Branch of Schmidt Institute of Physics of the Earth, Russian Academy of Sciences, Borok, Yaroslavl Region, Russia, ozotov@inbox.ru*

The essence of the cumulative effect of a round-the-world seismic echo is that the echo can serve as a trigger for a second tremors in the epicentral zone of the earthquake that gave rise to the echo. According to the classification of triggers, the round-the-world echo is an endogenous force mechanical additive trigger. The round-the-world echo excited by the main shock as a trigger for aftershocks has been studied in detail previously. In this work, the question of whether a foreshock can excite a round-the-world echo, which will turn out to be a trigger for the main shock, is posed and experimentally studied. In the course of the study, the classification of the so-called triads of earthquakes was considered. When studying one of the six types of triads, indirect signs were found that the answer to the question posed may be positive

### 1. Introduction

The cumulative effect of a round-the-world seismic echo was discovered 12 years ago in [1]. The essence of the effect is that a round-the-world echo can serve as a trigger for a repeated tremors in the epicentral zone of the earthquake that gave rise to the echo.

Let us indicate the location of this trigger in the general classification scheme for earthquake triggers. The classification outline is as follows. Triggers belong to one of two classes – endogenous and exogenous triggers. Sources of endogenous triggers are located under the earth's surface, and sources of exogenous ones are located above it. Each class is divided into three subclasses: force, thermal and chemical triggers. The subclass of force triggers contains two types: mechanical and electromagnetic. Each of these types has two varieties: additive and multiplicative triggers. The round-the-world echo is an endogenous force mechanical trigger.

Let us explain the difference between additive and multiplicative triggers using the example of aftershocks. The average dynamics of aftershocks is described by the logistic equation for the evolution of the source [2]

$$\frac{dn}{dt} = \gamma n \left(1 - \frac{n}{n_\infty}\right) + f(t).$$

Here $n$ is the average frequency of aftershocks, $\gamma$ and $n_\infty$ are phenomenological parameters that characterize the state of the source in a generalized form. The free term $f$ formally describes additive triggers in the form of external influences on rocks, inducing repeated tremors. A multiplicative trigger is an external influence leading to disturbance of the phenomenological parameters of the source. Apparently, the round-the-world echo should be classified as a type of additive trigger.

In this paper, we recall the excitation of aftershocks by round-the-world seismic echoes, consider the idea of the possible role of round-the-world echoes as a trigger for the main shock of an earthquake, and discuss the issue of periodic triggers of global seismicity.

## 2. Trigger excitation of aftershocks

When planning the study [1], our initial task was to search for an antipodean effect of the main shock of an earthquake. The point is as follows. The main shock excites a surface seismic wave, the front of which reaches the antipodal point relative to the earthquake epicenter after approximately 90 minutes. As the wave approaches the antipode point, the amplitude of seismic vibrations steadily increases. The question was posed: does a surface wave excite an aftershock with an epicenter at a point diametrically opposite to the epicenter of the main shock 90 minutes after the main shock?

The answer turned out to be the following: the antipodean effect of the main shock is not very clearly observed. The reason for the weakness of the antipodal effect was found after constructing a map on which points diametrically opposite the epicenters of strong earthquakes were plotted. It turned out that antipodal points are located primarily in aseismic regions of the planet. This circumstance prompted the idea to look not for the antipodean effect, but for the effect of a round-the-world seismic echo. As the echo approaches the epicenter of the main shock, the amplitude of the echo increases and there is reason to expect that in the epicentral zone, which is in a very nonequilibrium state, a strong aftershock will be excited 3 hours after the main shock.

Our expectation was fully justified. Initially, the effect of the round-the-world echo as a trigger of aftershocks was discovered through statistical analysis of earthquakes [1]. Then the effect was observed after catastrophic earthquakes in Sumatra (2004, $M = 9.2$) and Japan (2011, $M = 9.0$) [3]. A detailed study of the patterns of excitation of aftershocks by round-the-world echoes was carried out in [4] (see also review [5]).

We have no doubt that we are dealing with a real geophysical phenomenon. It is of interest for the physics of earthquakes, since the round-the-world echo, the properties of which are known quite well, produces a kind of probing of the earthquake source, relaxing after the main shock to a new state of equilibrium.

### 3. Classification of earthquake triads

Before the main shock of an earthquake, foreshocks are quite often, but not always, observed. Let $N_-$ and $N_+$ be the numbers of foreshocks and aftershocks for two equal periods of time before and after the main shock, respectively. Usually the inequality $N_- < N_+$ is observed (and even more so, as a rule $N_- \ll N_+$) Let us introduce a definition: a *classical triad* will be a unique trinity of foreshocks, the main shock and aftershocks at $N_- < N_+$ [6]. Sometimes foreshocks are completely absent even before very strong main shocks, i.e. $N_- = 0$. In this case, we will call the classical triad *shortened*, or *incomplete*.

One of the authors (O.D.) raised the question of searching for triads other than the classical triad [7]. We will call a *mirror* or *symmetric* triad in which the condition $N_- > N_+$ or $N_- = N_+$ is satisfied, respectively. Each of them can be either *full* or *shortened*.

Thus, we have a systematics of triads of earthquakes. All triads are divided into three classes, each of which contains two species. Let us present the taxonomy of triads in the form of Table 1.

Table 1. Classification of earthquake triads

| Triad T Foreshocks, Mainshock, Aftershocks | | | | | |
|---|---|---|---|---|---|
| Classical TC $N_- < N_+$ | | Mirror TM $N_- > N_+$ | | Symmetrical TS $N_- = N_+$ | |
| Complete TCC $N_- \neq 0$ | Incomplete TCI $N_- = 0$ | Complete TMC $N_+ \neq 0$ | Incomplete TMI $N_+ = 0$ | Complete TSC $N_\mp \neq 0$ | Incomplete TSI $N_- = N_+ = 0$ |

The cells of Table 1 indicate the full name, abbreviation and distinctive feature of each classification element. None of the six species is empty, as can be seen in Table 2. When compiling the table, the world catalog of earthquakes by U.S. Geological Survey (USGS/NEIC), 1973-2019 http://neic.usgs.gov/neis/epic/epic_global.html was processed. A total of 5259 triads were identified with main shock magnitudes of $M \geq 6$ and hypocenter depths of 0–250 km.

Table 2. Number of triads at main shock magnitudes $M \geq 6$.

| Species | TCC | TCI | TMC | TMI | TSC | TSI |
|---|---|---|---|---|---|---|
| Number | 332 | 2066 | 104 | 156 | 121 | 2480 |
| % | 6.3 | 39.3 | 2.0 | 3.0 | 2.3 | 47.1 |

The classification is useful in that it allows any triad to be classified into one of six well-defined types. But classification also has great heuristic significance. We will return to this issue in Section 5, but here we will only point out that the proposed classification allows us to immediately select the type of triads for experimental study of the question of whether a round-the-world echo excited by a foreshock could serve as a trigger for the main shock? It is natural to begin the pilot analysis by selecting TMI for the study despite the low likelihood of such events occurring.

## 4. Trigger excitation of the main shock

The discovery of a fairly large class of mirror triadsTM [7] prompted us to raise the question of whether the round-the-world echo of a sufficiently strong foreshock could serve as a trigger for the main shock, just as the round-the-world echo of the main shock serves as a trigger for a strong aftershock. Of course, it would be more natural to assume that foreshocks occurring in the immediate vicinity of the hypocenter of the future main shock are themselves direct triggers. However, the complex process of formation of the main rupture that initiates the main shock is not entirely clear to us. It cannot be ruled out that the direct impact of the foreshock has not yet triggered, while the appearance of a round-the-world echo completes the process of preparing the main rupture. One way or another, the question seemed interesting to us and we carried out a pilot analysis of incomplete mirror triads TMI.

We analyzed 156 TMI-type triads but did not find the expected trigger effect. We continued our search using triads like TMC, but also to no avail.

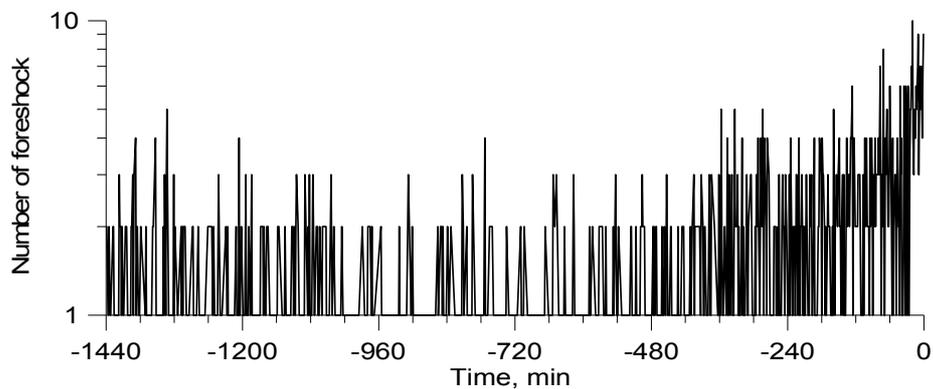

**Fig. 1**. Distribution of foreshocks in time according to the processing of 332 complete classical triads (see Table 2). Point 0 coincides with the moment of the main shock in each triad.

The result of the analysis of classical triads is of some interest. Figure 1 shows the distribution of foreshocks obtained by the epoch accumulation method. The time of the main shock in each of the triads of the TCC type, extracted from the USGS/NEIC world catalog, was chosen as a reference point. We do not find, as expected, a maximum frequency of foreshock occurrence in the vicinity of –180 min. (Recall that a similar maximum is clearly manifested in the activity of aftershocks [1, 3–6].) But what attracts attention is the rather sharp activation of foreshocks approximately 6 hours before the main shock.

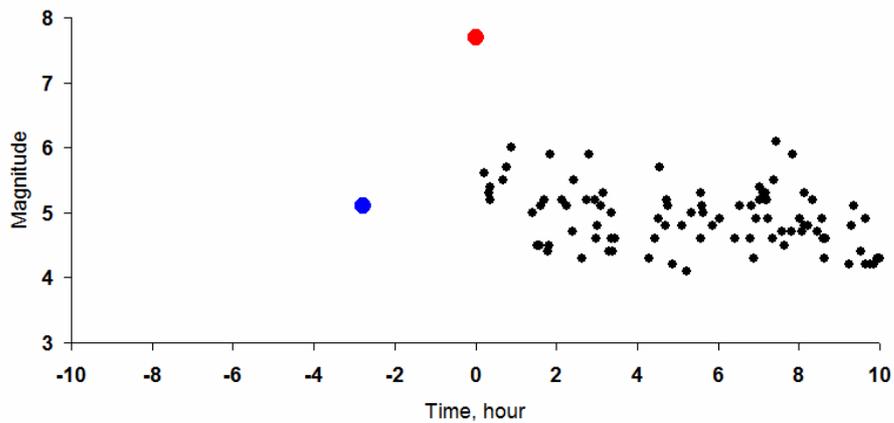

**Fig. 2.** An earthquake with a magnitude of *M*=7.7 occurred on 2006.07.17 8h19m26s at a depth of H=20 km. Epicenter coordinates: La = -9.28º , Lo = 107.42º. The foreshock is marked with a blue dot, the main shock with a red dot, and aftershocks with black dots.

So, statistical analysis of earthquakes registered in the world catalog USGS/NEIC allowed us to see the activation of foreshocks 6 hours before the main shock, but did not allow us to identify the expected trigger effect. In individual implementations of classical triads we sometimes see a seemingly clear manifestation of the effect. An example of this kind is shown in Figure 2. But it is quite obvious that individual examples are not enough to convince us of the detection of an effect.

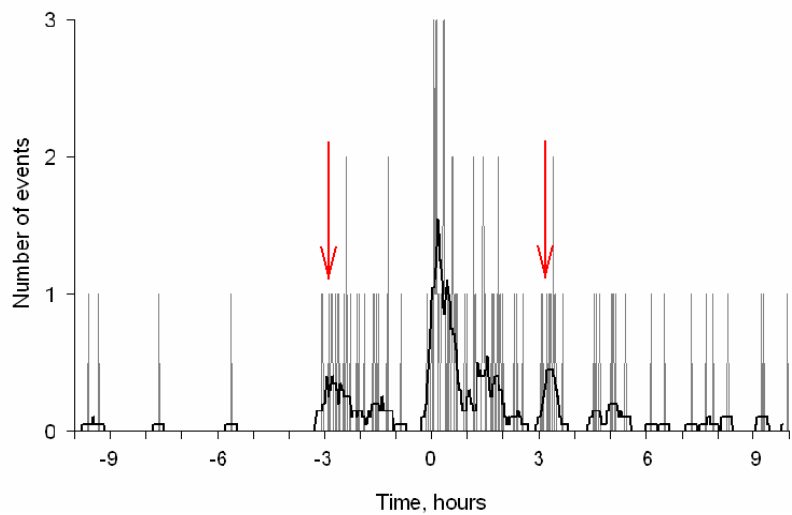

**Fig.3**. Dynamics of foreshocks and aftershocks with the magnitudes $4.5 \leq M < 5.5$ inside 3° epicentral zones of mainshocks with magnitudes $M \geq 5.5$, according to the regional catalog of earthquakes in Southern California (1983-2008).

Nevertheless, we believe that the possibilities for searching for the trigger effect of a round-the-world echo excited by a foreshock have not been exhausted. We are convinced of this by a preliminary analysis of the regional catalog of earthquakes in Southern California (http://www.data.scec.org). Figure 3 shows intensification of the foreshocks approximately 3 hours before the main attack. This indirectly indicates that the round-the-world seismic echo not only generates aftershocks, but can also be a trigger for the main shock.

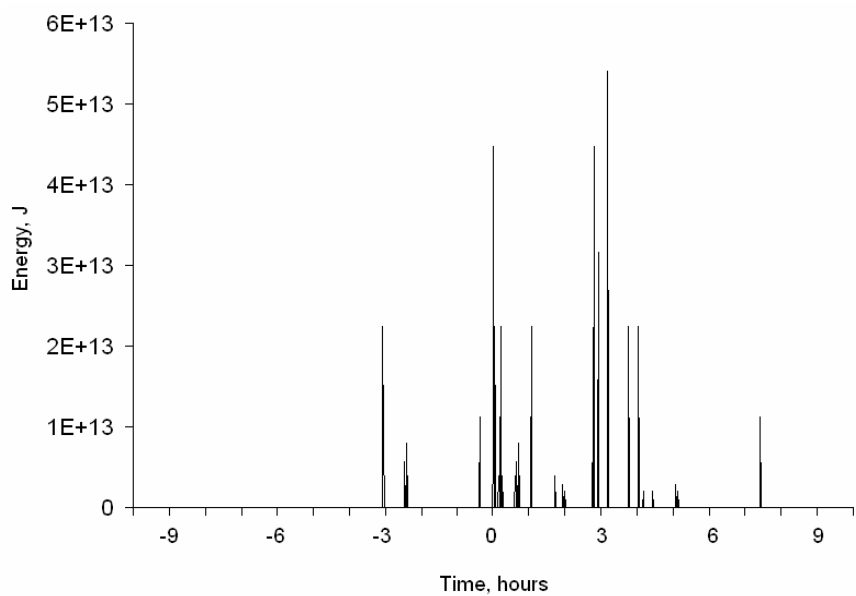

**Fig. 4.** Dynamics of foreshocks and aftershocks with the magnitudes 5≤ M <6 inside 3° epicentral zones of mainshocks with magnitudes M ≥ 6 according to the regional catalogs of earthquakes in Southern (1983-2008) and Northern (1968-2007) California.

Figure 4 shows the joint processing of Northern and Southern California earthquakes catalogs. Here we used the energy as a measure of foreshocks and aftershocks activity. (We did not consider here the energy of the mainshocks. In some events it reached several petajoules.) We see that the most strong foreshock is observed for 3 hours before the mainshock. This observation is an additional argument in favor of the idea is that the surface waves propagating outwards from the strong foreshock return back to the vicinity of the epicenter after having made a complete revolution around the Earth and induce there the mainshock.

## 5. Discussion

We will begin the discussion with the question of the classification of triads. The classification helped us select TMIs for pilot analysis to test the hypothesis that a round-the-world echo of foreshock could be a mainshock trigger. But in addition to the fact that our classification brought practical benefits, it turned out to be extremely valuable from a heuristic point of view. Indeed, in the process of systematizing triads, we identified a special and very common type of TSI consisting of isolated main shocks [7]. Such a main shock is not preceded by foreshocks, but aftershocks are not excited after it (see Table 2).

Next, the detection of the TM class allows us to refine the definition of the main shock. In addition to the well-known Bath inequality $\Delta M_+ \geq 1.1$ [9], one should add the $\Delta M_- \geq 0.5$ inequality [7]. Here $\Delta M_- = M_0 - \max\{M_-\}$ and $\Delta M_+ = M_0 - \max\{M_+\}$, where $M_-$, $M_0$, and $M_+$ are the magnitudes of the foreshocks, the main shock and the aftershocks, respectively.

Let us, however, move on to a discussion of earthquake triggers. Foreshocks, directly affecting the geological environment in the epicentral zone, stimulate, or more precisely accelerate the onset of the main rupture, i.e. are sort of triggers for the main shock. Taking into account the general patterns of rupture formation outlined by Rebetsky [8], it is natural to assume that, firstly, foreshocks lead to a redistribution of normal and tangential stresses in the source, and this, in turn, leads to the flow of fluid, which significantly reduces the level of critical stresses in fault zone. Secondly, seismic waves excited by foreshocks quite likely dynamically modify the material between the sides of faults, reducing its strength.

Let's borrow another idea from Rebetsky [8] and try to adapt it to our situation. We are talking about the qualitative difference in shear deformations in relatively weak and strong earthquakes. In the situation we are considering, it is apparently permissible to speak specifically about the foreshock and the main shock. It seems to us that when a foreshock is excited, the rupture along the fault surface occurs simultaneously, while the formation of the main rupture and the excitation of the main shock are a structured multi-stage process. The course of the process is affected by foreshocks. In this light, we can consider foreshocks as a kind of probing signals and monitor the corresponding response of the fault source zone. There is still a lot of work to be done in this direction, if it turns out to be productive. The class of mirror triads TM can be effectively used for an initial pilot analysis of a targeted problem.

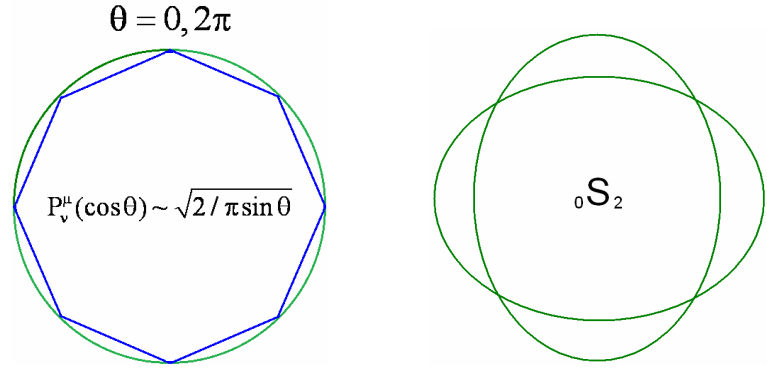

**Fig. 5**. Schematic pictures of the round-the-world seismic echo (left panel) and the spheroidal oscillations of the Earth (right panel). The resonant rays of the round-the-world echo formed by the surface and body waves (the smooth and broken lines, respectively). The angular dependence of the short-wave asymptotics of the associated Legendre polynomials, which are proportional to the amplitude of the oscillations, is shown in the central part of the image. The right panel shows a schematic representation of spheroidal oscillations.

There is no doubt that the main shock excites a round-the-world echo (see Figure 5), the cumulative effect of which quite often induces a strong aftershock. But in addition to the round-the-world echo, which can serve as a *pulse trigger* for an aftershock, the main shock excites free oscillations of the Earth, which modulate the activity of aftershocks, i.e. are a *periodic trigger* [1, 3, 5]. In particular, spheroidal oscillations of the Earth $_0S_2$ modulate seismicity with a period of 54 minutes (see Figure 5). At the same time, global seismicity experiences anthropogenic modulation with a period of 60 minutes [10, 11]. An idea arises to consider the lithosphere as a kind of mixer, the inputs of which are supplied with two signals, one at a frequency of $f_1 = 16.66$ mHz and the second at a frequency of $f_2 = 18.51$ mHz. If our idea is related to reality, then it is possible that at a beat frequency of $f_2 - f_1 = 1.86$ mHz one can, in principle, observe a modulation of global seismicity. The beating period is approximately 9 hours. On this fantastic note, we will conclude our discussion of earthquake triggers.

## 6. Summary

1. The phenomenon of the triggering effect of a round-the-world seismic echo excited by the main shock on the earthquake source should be considered established quite reliably. The phenomenon was discovered by the authors 12 years ago. It has stood the test of time. The increase in aftershock activity 3 hours after the main shock has been confirmed many times through the study of a large volume of observational data.

2. There is no doubt that the excitation of foreshocks directly stimulates the formation of a main rupture at the source, which manifests itself in the form of the main shock of an earthquake. Of interest is the possible difference in the mechanisms of the main shocks in complete and incomplete classical triads of earthquakes.

3. We consider the hypothesis that round-the-world echoes excited by foreshocks can from time to time serve as triggers for the main shocks of earthquakes to be very plausible. A number of indirect signs support our hypothesis. Unusual events of this kind occur infrequently. Their search and careful study is of extremely great importance for the physics of earthquakes.

4. In the process of this research, a taxonomy of earthquake triads was developed. The triads are divided into three classes and together contain six species. We have indicated very well-defined criteria for identifying species of triads.

*Acknowledgements*. We express our deep gratitude to B.I. Klain and A.D. Zavyalov for his attention to the work and meaningful discussions. We are grateful to the staff of the US Geological Survey for providing earthquake catalogs. The work was carried out in accordance with the plan of state assignments of the IPE RAS.